\documentstyle[12pt,aps]{revtex}

\makeatletter                    
\@addtoreset{equation}{section}  
\makeatother                     

\begin{document}
\title{Wave Propagation in Stochastic Spacetimes: Localization,
Amplification and Particle Creation}
\author{B. L. Hu and K. Shiokawa \\
{\small  Department of Physics, University of Maryland, College Park, 
MD 20742, USA}\\
 \small{\it(umdpp 97-125, submitted to Phys. Rev. D on July 29, 1997)} }
\maketitle
\begin{abstract} 
We study novel effects associated with electromagnetic wave
propagation in a Robertson-Walker universe and the Schwarzschild spacetime
with a small amount of metric stochasticity.
We
find that localization of electromagnetic waves occurs
in a Robertson-Walker universe with time-independent metric stochasticity,
while time-dependent metric stochasticity
induces exponential instability in the particle production rate.
For the Schwarzschild metric, time-independent randomness
can decrease the total luminosity  of Hawking radiation
due to multiple scattering of waves
outside the black hole and
gives rise to event horizon fluctuations and thus
fluctuations in the Hawking temperature.
\end{abstract}

\newpage
\section{Introduction}

Wave propagation and localization \cite{Anderson58}
in a random media has been studied extensively
for the last two decades \cite{Frish68,Ishimaru97,WPRM}.
It is known that on the mesoscopic scale, classical wave propagation in a
random media can be treated in a similar way as electron transport in a random
potential \cite{WPRM}. One expects to see the diffusive, localization properties
of classical waves similar to electrons moving in the presence of impurities 
\cite{LeeRama,rmedia}.
Wave propagation in curved spacetime is an important topic both in
general relativity \cite{MTW} and in semiclassical gravity theory \cite{BirDav}.
Classical scalar, electromagnetic and gravitational waves
in Friedmann-Robertson-Walker (FRW) Universe probe into 
the state of the universe and manifest in basic cosmological processes 
such as structure
and defect formation, while that in the Schwarzschild and Kerr
spacetimes depict high energy astrophysical processes in black holes.
The second-quantized version in terms of quantum fields gives rise to
cosmological particle creation  \cite{cpc} and 
Hawking radiation \cite{Hawking74}
which are important processes in the early universe and black holes collapse.
Recent progress in studying Planck energy processes, especially the
backreaction effect of quantum fields in curved spacetime, underscores 
the importance of including fluctuations in particle creation 
\cite{HuPhysica,CH94} and the
associated energy momentum tensor of quantum fields \cite{KuoFor,PhiHu,Wald},
and fluctuations and dissipation in the dynamics of spacetime
\cite{HM3,fdrsc,CamVer96,CCV}.

The program devoted  to
quantum matter field and classical background spacetime
with metric fluctuations mentioned above is rather involved,
 because it requires the calculation of four point functions \cite{HPR}
  and demands a self-consistent solution \cite{CHR,HRS}. In this paper,
as a useful parallel,  we attempt to address an easier problem, that of
wave propagation in a stochastic spacetime. It is  designed to highlight
the effect of fluctuations in a background metric, 
while not demanding an explanation
of their source, or their mutual influence. 
Stochastic components in the metric can be induced by primordial gravitational
waves, topological defects in the sub-Planckian scale,
or intrisic metric fluctuations of background spacetimes at the Planck scale.
Their detection and analysis can provide valuable information about the state
of the early universe and black holes. As distinct from the self-consistent
treatment which is necessary for Planck scale processes, wave propagation in
curved spacetimes with metric stochasticity is a test-field treatment.
It is nonetheless still a useful probe for fluctuations in sub-Planckian
processes (such as GUT scale phase transition), which could have left
important imprints on the observable universe \cite{Allen96}.

In this paper,
we wish to study novel effects associated with electromagnetic waves
propagation in the Friedmann-Robertson-Walker (FRW) universes 
and the Schwarzschild spacetime
with a small amount of metric stochasticity. Here we employ a useful observation
to link up with the more familiar subject of wave propagation in random
media studied extensively in condensed matter 
and mesoscopic physics \cite{LeeRama,rmedia}.
We first show the formal equivalence of the wave equations in curved spacetimes
with wave propagation in a media in flat space and
identify how the metric components appear in the permittivity function
(or refractive index) of the media \cite{Mashhoon1}.
Then we introduce metric fluctuations as a stochastic
component in the permittivity function and study wave propagation in a curved
spacetime with metric stochasticity as if it were in a random media.
In a spherically symmetric spacetime the wave equation
for the radial part can be written in the form of a Schr\"{o}dinger equation
in one dimension. The effect of the curvature of spacetime appears
in the potential term in the equation. Once the wave equation is reduced
to the parametric form with a stochastic component, the familiar methods
used in quantum field theory in curved spacetime and insights
accumulated in mesoscopic physics can work in each other's advantage.
We analyse in detail the cases with time-independent and 
time-dependent metric stochasticity and find that 
localization of electromagnetic waves occurs in a metric 
with time-independent fluctuations.
 In cases where there are time-dependent randomness
in the metric, exponential instability in the particle production rate
occurs. These are new effects due exclusively to the presence
of metric fluctuations in the background spacetimes.
For the Schwarzschild metric,
time-independent randomness outside the horizon
will decrease the total luminosity  of Hawking radiation due to multiple
scattering of waves.
if the randomness reaches the horizon, it contributes to the fluctuations of
Hawking temperature.
Time-dependent stochasticity is a more complicated matter
which requires a self-consistent analysis of the interaction between spacetime
and the waves or fields as encoded in the 
fluctuation-dissipation relations \cite{CHR,HRS}. 
This is to be investigated later as a part of the
stochastic backreaction problem mentioned above.

The paper is organized as follows:
In Sec. 2, we show how an electromagnetic wave propagating in curved
spacetime can be related to that in flat space but with a refractive index
depending on the metric components \cite{Mashhoon1}. This section is meant
to be a short cut for readers not too familar with curved spacetime physics
to see the correspondence with wave phenonema. (Readers familiar with it can
skip to Sec. 3).  In Section 3, using the methods developed in
Sec. 2 we study wave propagation in curved spacetime
where the metric has a stochastic component. 
We use the Friedmann-Robertson-Walker universe as an example 
to show that if the stochastic component of the metric is independent of
time,  and for a  sufficiently smooth randomness
the Maxwell equation has the same form 
as a conformally coupled scalar wave equation.
If the randomness appears only in the radial direction,
the radial wave equation has the same form as an one dimensional Schr\"{o}dinger
equation in a random potential. Because the electromagnetic
wave  equation is conformally invariant,  
it bears the  same form as in flat space. With a stochastic metric component
we see that wave localizes in space. In Sec. 4, we study the case of
time-dependent (but space-independent) stochasticity in the metric, and
show that  parametric amplification takes place giving rise to cosmological
particle creation. However, because of the metric fluctuations
the rate of this amplification increases exponentially in time.
The fluctuations in the particle creation rate is also discussed.
In Sec. 5, we study wave propagation in a Schwarzschild spacetime with
time-independent fluctuations.
We show that this can
decrease  the total luminosity of Hawking radiation due to multiple scattering
of waves  outside the black hole.
 Fluctuations which involve event horizon
 will give rise to fluctuations in the Hawking temperature.
 In Sec. 6 we summarize our findings and end with a short discussion.

\section{Classical Electromagnetic Waves in Curved Spacetime}

Maxwell's equations  for an electromagnetic field tensor $F_{\mu \nu}$ 
in a gravitational field with metric $g_{\mu \nu}$ are given by
\begin{equation}
	F^{\mu \nu}_{\nu} = 0 ~~~
	F_{\mu \nu ; \sigma} + F_{\nu \sigma ; \mu} + F_{\sigma \mu ; \nu}= 0
\end{equation}
where semicolons denote covariant derivatives with respect to $g_{\mu \nu}$.
These equations can be cast in a form for waves propagating in a
permeable media in flat space \cite{Mashhoon1}. The correspondence
between  $E, H$ (the electric field and magnetic induction) and
$D, B$ (the electric displacement and magnetic field) respectively is
given by
\begin{eqnarray}
	D_{i} &=& \epsilon_{ik} E_{k} - (G \times H)_{i}  \nonumber \\
	B_{i} &=& \mu_{ik} H_{k} - (G \times E)_{i}
\end{eqnarray}
where the dielectric and permitivity functions are 
\begin{eqnarray}   
   \epsilon_{ik} &=& \mu_{ik} = - (-g)^{1/2} \frac{g^{ik}}{g_{00}} \nonumber \\
	 G_{i}   &=& - \frac{g^{0i}}{g_{00}} .
    \label{B}
\end{eqnarray}

\subsection{ Friedmann-Robertson-Walker Universe}

Let us now consider the Friedmann-Robertson-Walker (FRW) spacetimes 
with line elements
\begin{eqnarray}
-ds^2 &=& -dt^2 + a(t)^2 R^2 [d \chi^2 +
s^2(\chi)
%
(d \theta^2 + \sin^2 \theta d \phi^2)]
\end{eqnarray}
%
where $s(\chi) = \sin \chi, \chi, \sinh \chi$ 
correspond to closed, flat, and open cases
respectively.
 Here $a(t)$ is the scale factor and $R$ is the radius at time
$t_0$ where  $a(t_0) = 1$.

Using Cartesian coodinates, 
\begin{eqnarray}
dt  &=& a(t) d \eta                                \nonumber \\
x^1 &=& 2 R t(\chi/2) \sin{\theta} \cos{\phi}   \nonumber \\
x^2 &=& 2 R t(\chi/2) \sin{\theta} \sin{\phi}   \nonumber \\
x^3 &=& 2 R t(\chi/2) \cos{\theta}
\end{eqnarray}
where 
$\eta \equiv \int dt/a$ is the conformal time, and
$t(\chi/2) = \tan(\chi/2), \chi/2, \tanh(\chi/2)$ 
correspond to  the closed, flat, and open cases
respectively.
We can then write the FRW  line element in the form
\begin{eqnarray}
-ds^2 &=& a^2 [ -d \eta^2 + f^2(\rho) (\delta_{ij} dx^i dx^j) ]
\end{eqnarray}
where
\begin{eqnarray}
\rho &=& [ \sum_{i} (x^i)^2 ]^{1/2}       \nonumber \\
	  &=& 2 R t(\chi/2)
\end{eqnarray}
and
\begin{eqnarray}
f(\rho) = 
\left(
\begin{array}{c}
\frac{1}{1 + \rho^2 / 4 R^2}  \nonumber \\
1                             \nonumber \\
\frac{1}{1 - \rho^2 / 4 R^2}
\end{array}
\right)
\end{eqnarray}
where the column elements correspond to closed, flat, and open cases
respectively.

In this isotropic form of the metric, the Maxwell equations
 are given by
\begin{eqnarray}
 \nabla \times \vec{E} &=& - \frac{d}{d \eta} (f(\rho) \vec{H})  \nonumber \\
 \nabla \times \vec{H} &=& \frac{d}{d \eta} (f(\rho) \vec{E})  \nonumber \\
 \nabla \cdot (f \vec{E}) &=& 0    \nonumber  \\
 \nabla \cdot (f \vec{H}) &=& 0              \label{f2}
\end{eqnarray}
where one can identify the dielectric and permittivity functions as
\begin{eqnarray}
	\epsilon_{ik} &=& \mu_{ik} = f(\rho) \delta_{ik}   \nonumber \\
			G_{i}   &=& 0 .                  \label{A}
\end{eqnarray}

 Assuming a harmonic (conformal) time dependence of the solution
 $e^{- i \omega \eta}$, we can write (\ref{f2}) as
\begin{eqnarray}
 \nabla \times \vec{E} &=& i \omega f(\rho) \vec{H}   \nonumber \\
 \nabla \times \vec{H} &=& - i \omega f(\rho) \vec{E} \nonumber \\
 \nabla \cdot (f \vec{E}) &=& 0		   \nonumber \\
 \nabla \cdot (f \vec{H}) &=& 0  .         \label{D}
\end{eqnarray}
The symmetry in the relation (\ref{D}) allows us to write  this in
the more compact form
\begin{eqnarray}
 \nabla \times \vec{F} &=& \omega f(\rho) \vec{F}  \nonumber \\
 \nabla \cdot (f \vec{F}) &=& 0
       \label{f4}
\end{eqnarray}
where  $\vec{F} \equiv \vec{E} + i \vec{H}$.


For the FRW metric with spherical symmetry  the solutions can be expressed
in terms of the vector spherical harmonics
 $\vec{Y}_{lm}(\theta,\phi)$ obtained by operating on
 the scalar spherical harmonics $Y_{lm} (\theta, \phi)$ with the invariant 
operator $-i r \cdot \nabla$ as
\begin{eqnarray}
 \vec{E}
 &=& \sum_{l,m} [  \frac{i}{\omega f}
         A^{E}_{lm} \nabla \times g^{E}_{l}(\rho) \vec{Y}_{lm}
        +A^{M}_{lm} g^{M}_{l}(\rho) \vec{Y}_{lm}   ]     \nonumber \\
 \vec{H}
 &=& \sum_{l,m} [
 A^{E}_{lm} g^{E}_{l}(\rho) \vec{Y}_{lm}
 - \frac{i}{\omega f}
 A^{M}_{lm} \nabla \times g^{M}_{l}(\rho) \vec{Y}_{lm}   ]
\end{eqnarray}
where $g^{E,M}_{l}(\rho)$ are functions of radial distance only
 and $ A^{E,M}_{lm} $ are the coefficients. Superscripts $E$ and $M$ denote
  the electric and magnetic
multipole field components, respectively.

The field $ \vec{F} $ is given by
\begin{eqnarray}
 \vec{F}
 &=& \sum_{l,m}
  [ i A^{E}_{lm}
  (
	 \frac{1}{\omega f}
         \nabla \times g^{E}_{l}(\rho) \vec{Y}_{lm}
         +             g^{E}_{l}(\rho) \vec{Y}_{lm}
  )
   \nonumber \\
 &+&     A^{M}_{lm}
  (
         g^{M}_{l}(\rho) \vec{Y}_{lm}
	 + \frac{1}{\omega f}
                \nabla \times g^{M}_{l}(\rho) \vec{Y}_{lm}
  )
  ]   .
\end{eqnarray}
One can identify the $(l,m)$ component of a magnetic multipole field 
 $ \vec{F}^{M}_{lm} $ as
\begin{eqnarray}
 \vec{F}^{M}_{lm}
 &=&
         g^{M}_{l}(\rho) \vec{Y}_{lm}
	 + \frac{1}{\omega f}
                \nabla \times g^{M}_{l}(\rho) \vec{Y}_{lm}   .
\end{eqnarray}
%

%
%
Writing the magnetic multipole scalar field function
\begin{eqnarray}
\psi^{M}(\chi) = R \rho(\chi) g^{M}_{l}(\rho(\chi)),
\end{eqnarray}
we obtain the radial equation for $\psi^{M}$ as
\begin{eqnarray}
\frac{d^2 \psi^{M}(\chi)}{d \chi^2} + (n^2 - U(\chi)) \psi^{M}(\chi) = 0
\label{f99}
\end{eqnarray}
where $n = R \omega$ and
\begin{eqnarray}
 U(\chi) = \frac{j(j+1)}{s^2(\chi)}      .
\end{eqnarray}
%
%
%
%
%

\subsection{Schwarzschild Spacetime}

The line element for the Schwarzschild spacetime is given by
\begin{eqnarray}
-ds^2 &=& -( 1 - \frac{2M}{r} ) dt^2 + ( 1 - \frac{2M}{r} )^{-1}
	d r^2 + r^2 [d \theta^2 + \sin^2 \theta d \phi^2]   .
	 \label{f12}
\end{eqnarray}
If we define $r \equiv \rho + M + M^2/4 \rho $, we can write (\ref{f25})
as
\begin{eqnarray}
-ds^2 &=& - f_1(\rho) dt^2 + f_2^2(\rho)
  [
	d \rho^2 + \rho^2 d \theta^2 + \rho^2 \sin^2 \theta d \phi^2
   ]
   	 \label{f26}
\end{eqnarray}
where $f_1(\rho)=  1 - \frac{2M}{r}$ and $f_2(\rho)= r/\rho$.
Furthermore, a spatial coordinate transformation
\begin{eqnarray}
x^1 &=& \rho \sin{\theta} \cos{\phi}   \nonumber \\
x^2 &=& \rho \sin{\theta} \sin{\phi}   \nonumber \\
x^3 &=& \rho \cos{\theta}
\end{eqnarray}
brings (\ref{f26}) into a Cartesian form
\begin{eqnarray}
-ds^2 &=& - f_1(\rho) dt^2 + f_2^2(\rho)
      [	(d x^1)^2 + (d x^2)^2 + (d x^3)^2 ]    .
   	 \label{f27}
\end{eqnarray}
Considering this spacetime acting as a medium, the corresponding
 dielectric and permittivity functions are expressed in the same
 form as in (\ref{A}) with
\begin{eqnarray}
	 f(\rho) = \frac{ f_2(\rho) }{ (f_1(\rho))^{1/2} }    .
	 \label{f28}
\end{eqnarray}

In such a medium, the radial equation for the magnetic
 multipole field is \cite{Mashhoon2}
\begin{eqnarray}
\frac{d^2 \psi^{M}(r*)}{d r*^2} + [ \omega^2 - U(r*) ] \psi^{M}(r*) = 0
  \label{f30}
\end{eqnarray}
where $r* = r + 2M \log (r/2M - 1)$
is the Regge-Wheeler (turtoise) coordinate,
and the potential term is given by
\begin{eqnarray}
 U(r*) = \frac{ j(j+1) }{ r^2(r*) }  [ 1 - \frac{2M}{r(r*)} ].
\end{eqnarray}

\section{Wave Propagation in Curved Spacetime with Metric Stochasticity}

Consider the metric  as containing  an averaged (deterministic) part  
$g_{\mu \nu}^{(0)}$ and  a small fluctuating  (stochastic) part  $h_{\mu \nu}$
\begin{eqnarray}
 g_{\mu \nu} = g_{\mu \nu}^{(0)} + h_{\mu \nu}
\end{eqnarray}
then from (\ref{B})
\begin{eqnarray}
	\epsilon_{ik} &=&
	\mu_{ik} = - \frac{(-g^{(0)})^{1/2}}{g^{(0)}_{00}} g^{(0) ik}
			+ \frac{(-g^{(0)})^{1/2}}{g^{(0)}_{00}}
			( h^{ij} - \frac{1}{2} h  g^{(0) ik}  )
	  \nonumber \\
			&=&
	  \epsilon^{(0)}_{ik} + \delta \epsilon_{ik}
\end{eqnarray}
where we  have used the synchronous  $h_{00}=0$ gauge \cite{Lifshitz46}.
 The traceless $h \equiv \sum_{i=1}^{3} h_{i}^{i}=0$ gauge further simplifies 
the expression and we obtain the correlation function of the fluctuating
 part of the refractive index in terms of the metric fluctuations as

\begin{eqnarray}
  \langle \delta \epsilon_{ij}(x) \delta \epsilon_{kl}(x') \rangle   &=&
	\frac{ ( -g^{(0)}(x) )^{1/2} ( -g^{(0)}(x') )^{1/2} }
			  { g^{(0)}_{00}(x)  g^{(0)}_{00}(x')  }
  \langle h^{ij}(x) h^{kl}(x') \rangle.
\end{eqnarray}
For flat spacetime, it has the simple form
\begin{eqnarray}
  \langle \delta \epsilon_{ij}(x) \delta \epsilon_{kl}(x') \rangle   &=&
	 \langle h^{ij}(x) h^{kl}(x') \rangle.
\end{eqnarray}
We see metric fluctuations can thus  be represented as fluctuations
in the optical index.

\subsection{Time Independent Metric Stochasticity}

In the rest of this section,
we consider cases in which the permitivity function contains only
spatial disorder.

 If we assume the background metric as well as the stochasticity
 are both isotropic, we can write the corresponding refractive index as
\begin{eqnarray}
	\epsilon_{ik} &=& \mu_{ik} = f(\vec{x}) \delta_{ik}   \nonumber \\
  	      G_{i}   &=& 0  
                 \label{f109}
\end{eqnarray}
where $f(\vec{x})$ is a random variable.

Assuming a harmonic time dependence of the field with frequency
$\omega$,
Eq. (\ref{f4}) with source term $\vec{J}$ becomes  
\begin{eqnarray}
 (   \vec{L} -  \omega f(\vec{x}) )   \vec{F}	 =   \vec{J}  \nonumber \\
						\label{C}
\end{eqnarray}
where $ \vec{L} \equiv \nabla \times $ .
Define Green's function $g(x,x')$ as
\begin{eqnarray}
 (   \vec{L} -  \omega f(\vec{x})  )  g(x,x')  =  \delta(x - x').
						 \label{f111}
\end{eqnarray}
Then the disorder-averaged Green's function is
\begin{eqnarray}
  G(x,x') &\equiv&  \langle g(x,x') \rangle              \nonumber \\
& = &     \langle (   \vec{L} -  \omega f(\vec{x})  )^{-1}  \rangle .
					  \label{f112}
\end{eqnarray}
If we define the mean field Green's function $G^{(0)}(x,x')$ as
\begin{eqnarray}
  G^{(0)}(x,x') &\equiv&   
		(
	 \vec{L} -  \omega \langle f( \vec{x}) \rangle   \label{f113}
                )^{-1}
\end{eqnarray}
we obtain the Lippman-Schwinger equation
\begin{eqnarray}
	G^{-1}(x,x')   =   G^{(0) -1}(x,x') + \omega \Sigma(x,x')
						  \label{f114}
\end{eqnarray}
where $\Sigma(x,x')$ is the self-energy.
Also averaging (\ref{C}), we have
\begin{eqnarray}
  \langle  \vec{L} \vec{F} \rangle  -  \omega \langle f(\vec{x}) \vec{F} 
\rangle	 =   \vec{J}  .
						\label{f115}
\end{eqnarray}
Define $ \vec{\cal F} \equiv f \vec{F} = \vec{D} + i \vec{B}$,
then from (\ref{f115})
\begin{eqnarray}
 \omega \langle \vec{\cal F} \rangle &=&
\langle  \vec{L} \vec{F} \rangle  -  \vec{J}   \nonumber \\
						  &=&
	 \int dx'
( G^{(0) -1}(x,x') + \omega \Sigma(x,x') - \vec{L} )  \langle \vec{F} \rangle
						 \label{f116}
\end{eqnarray}
\begin{eqnarray}
 \langle \vec{\cal F}(x) \rangle &=&
	 \int dx'
(\langle  f(\vec{x}) \rangle  \delta(x - x') - \Sigma(x,x') ) \langle \vec{F}
(x') \rangle           .
	 					 \label{f117}
\end{eqnarray}
We see that the complex electromagnetic
displacement vector has a nonlocal dependence on the
media due to the self-energy term.
We can thus develop a formalism making use of
techniques from quantum field theory to
study the effect of dissipation, higher order correlation,
screening, etc.  (see e.g., \cite{Balian95}).
This is a subject for future investigation.

Further simplification arises in cases when the polarization effects are
negligible.
From (\ref{D}), we obtain
\begin{eqnarray}
 \nabla^2 \vec{E} + \omega^2 f^2(\vec{x}) \vec{E}
 +   \nabla \log f(\vec{x}) \times \nabla \times \vec{E}
 +   \nabla ( ( \vec{E} \cdot \nabla ) \log f(\vec{x}) )
 = 0              .
 \label{f41}
\end{eqnarray}
Under the conditions that the inhomogeneity and disorder
in $f$ are smooth, namely,
 $ \omega R >> 1 $ and $ \omega \lambda >> 1 $,
  where $\lambda$ is the characteristic length scale of the
 disorder, the third and the fourth term representing polarization
  effects are negligible.
  (\ref{f41}) then reduces to
\begin{eqnarray}
 \nabla^2 \vec{E} + \omega^2 f^2(\vec{x}) \vec{E}
 = 0   .
\end{eqnarray}
Then each component satisfies the scalar wave equation
\begin{eqnarray}
 \nabla^2 \phi + \omega^2 f^2(\vec{x}) \phi
 = 0    .
 \label{f50}
\end{eqnarray}
Now consider the conformal type of stochasticity
only in the spatial part of the
metric of the form
\footnote{
The conformal invariance of Maxwell's equations implies that
the conformal fluctuations in the spacetime metric can be factorized away, 
and hence they do not give rise to particle creation \cite{Parker69}.
}
\begin{eqnarray}
  f(\vec{x})
   &=&
	 f(\rho) e^{g \sigma(\vec{x})},
    \label{f9}
\end{eqnarray}
where $\sigma(\vec{x})$ is a stochastic function and
 $g$ is a small parameter.
Expanding it with respect to $g$ as

\begin{equation}
	 f(\vec{x}) =  f(\rho)(1 + g \sigma(\vec{x})),
\end{equation}
(\ref{f50}) becomes
\begin{eqnarray}
\nabla^2 \phi + \omega^2 f^2(\rho) \phi
			+ 2 g \omega^2 f^2(\rho) \sigma(\vec{x}) \phi
 = 0                        \label{K}
\end{eqnarray}
%
%
%
%
%
In  a flat FRW spacetime $f(\rho) = 1$, or,
for $\rho << R$ in closed and open cases, $f(\rho) \sim 1$.
In this region, all three cases are described by the equation in flat spacetime
\begin{eqnarray}
\nabla^2 \phi + \omega^2 \phi + 2 g \omega^2 \sigma \phi
 = 0  .
\end{eqnarray}
The corresponding equation for the Green function is
\begin{eqnarray}
( \nabla^2 + \omega^2 + 2 g \omega^2 \sigma )
 G(\vec{x},\vec{x'},\omega)
 = \delta(\vec{x} - \vec{x'})   .
\end{eqnarray}
The disorder-averaged Green function
 $\langle G(\vec{k},\vec{k'},\omega)\rangle = G(\vec{k},\omega) \delta_{kk'}$
 is known to have a form
\begin{eqnarray}
G(\vec{k},\omega)  =
\frac{1}{\omega^2 - \vec{k}^2 + i \frac{\omega}{l} }
\end{eqnarray}
in the leading order in $(\omega l)^{-1}$,where $l$ 
is an elastic mean free path.
For a Gaussian white noise (local) random variable $\sigma(\vec{x})$
, $l = 2 \pi/  g D \omega^4$ where $D$ is the strength of disorder
defined by $\langle \sigma(\vec{x}) \sigma(\vec{x'}) \rangle
= D \delta(\vec{x} - \vec{x})$.

We also define the intensity
\begin{eqnarray}
 I(\vec{x}) = \langle \phi^{*}(\vec{x}) \phi(\vec{x'}) \rangle
\end{eqnarray}
and the intensity-intensity correlation function
\begin{eqnarray}
C( \vec{x},\vec{x'} )
=
\langle   \phi^{*}(\vec{x}) \phi(\vec{x}) \phi^{*}(\vec{x'}) \phi(\vec{x'})
 \rangle_C
\end{eqnarray}
where $C$ denotes the cumulant.
%
%
%
%
As shown in \cite{Shapiro86},
we can evaluate the  disorder-averaged intensity as
\begin{eqnarray}
\langle I(\vec{x})\rangle
= \frac{3}{16 \pi^2 l |\vec{x}|}   .
\end{eqnarray}
%

%
%
%
%
%
%
%
%

If we confine the disorder in a slab of thickness $L$,  and cross-sectional
area $A$, then for $r \equiv |\vec{x}-\vec{x'}| < l$,
\begin{eqnarray}
C(\vec{x},\vec{x'})
=
\langle I(\vec{x}) \rangle \langle I(\vec{x'}) \rangle
( \frac{\sin \omega r}{  \omega r} )^2
e^{-   \frac{r}{ l }  }.
\end{eqnarray}
For $r > l$, diffusion modes interact  with each other and
give rise to a long range correlation with  a power law decay \cite{SteCwi}.
In the Schr\"{o}dinger equation analogy,
this  corresponds to conductance fluctuations in metals \cite{CondF}.
%
%
%
%
%
%
%
%

\subsection{Radial Disorder -- Localization}

Here we also start from the conformal type of metric stochasticity but
 assume that the stochasticity is only in the radial direction
\begin{equation}
	 f(x) = \frac{1}{1 + t^2 (\chi/2)} e^{g \sigma(\chi)}
\end{equation}
where $g$ is a small parameter
( Such kind of randomness may arise from averaging out more
  inhomogeneous types
  of randomness in a small spatial region due to the spherically
  symmetric nature of the underlying curved spacetime).

If $\sigma(\chi)$ varies slowly in space ($\sigma' << \sigma$ ), 
the wave equation
in the radial direction (\ref{f99}) is replaced by
\begin{equation}
	 \frac{d^2 \psi}{d \chi^2} +
	 [ n^2(1 + g \sigma)^2 - \frac{j(j+1)}{ s^2 (\chi) }] \psi = 0
  \label{f5}
\end{equation}
where we simply use $\psi$ for $\psi^{M}$.
This has the same form as a
scalar wave equation in a media with a random refractive index.

For the closed FRW  universe,
the angular-momentum induced potential term in (\ref{f5}) is 
a single well potential which becomes infinitely high at the origin and
$\chi = \pi$.
Then the wave is necessarily localized around the bottom of the well
in this coordinate. On the other hand, for the flat and 
open  FRW universe cases,
the potential term becomes asymptotically flat as the radius becomes infinite.
In these cases, the stochastic term plays
a significant role in the transport property
of the wave.
If we expand the second term in (\ref{f5}) into $n^2 + 2 g n^2 \sigma $
around the perturbation parameter $g$
and note the positivity of the original second term,
we see (\ref{f5}) describes a Schr\"{o}dinger equation with a fixed
positive energy. As such , we can write (\ref{f5}) as
\begin{equation}
	- \frac{d^2 \psi}{d \chi^2} + V(\chi) \psi = E \psi
\label{f6}
\end{equation}
where $E \equiv n^2$ and $V(\chi) \equiv 
- 2 g n^2 \sigma + [j(j+1) / s^2 (\chi)]$.

For the flat or open universes, the potential barrier vanishes 
in the asymptotic limit.
The property of the eigenfunction in this case is well-studied from
the context of electron transport \cite{WPRM,LeeRama,rmedia}.
It is known that in one dimension for any value of $E$
, the eigenstate of Eq. (\ref{f6}) localizes
\cite{Anderson58,WPRM}.
Namely, all eigenfunctions decay exponentially with rate given by
 the Lyapunov exponent  [see (\ref{f7}) below].
If the correlation radius $\chi_c$ of the fluctuating part of the potential
 $V(\chi)$ is smaller than
the "wavelength"
\footnote
{
The "energy" $E = R^2 \omega^2$ in (\ref{f6}) 
is dimensionless and so is the angular
variable $\chi$.
}
($\chi_c << n^{-1}$),
the potential can be considered as the white noise type
\begin{equation}
 \langle V(\chi) V(\chi')\rangle =
 4 g^2 n^4 \langle \sigma(\chi) \sigma(\chi')\rangle =
 4 g^2 R^4 \omega^4 D \delta(\chi - \chi')
 \label{f40}
\end{equation}
where we have assumed a delta function type of correlation 
with the strength $D$
 for the random potential
$\sigma(\chi)$.
\footnote
{
Hereafter we absorb $g$ in the definition of $D$
}

The cumulative density of states
$N(\omega)$ has the following form
 for the white noise potential with correlation given in (\ref{f40})
  \cite{Halperin65}.
\begin{eqnarray}
 N(\omega) = \frac{(2 R^2 \omega^2 D)^{1/3} }{\pi^2}
 \frac{1}{   Ai^2(- (R \omega / 2 D)^{2/3} )
			  + Bi^2(- (R \omega / 2 D)^{2/3}) }
\end{eqnarray}
 where $Ai$ and $Bi$ are two independent Airy functions which satisfy
 the equation $y'' - xy = 0$.

The asymptotic behavior of the
solution shows an exponential growth characterized by the Lyapunov exponent
$\lambda_{\omega}$ defined by \cite{rmedia}
\begin{equation}
 \lambda_{\omega} = \lim_{\chi \rightarrow \infty}
 \lambda_{\omega}(\chi) = \frac{1}{2 L_{loc}(\omega)}
\label{f7}
\end{equation}
where $L_{loc}(\omega)$ is the localization length given below
and the wave function grows as
\begin{equation}
 \psi(\chi) \rightarrow e^{\lambda_{\omega}(\chi) \chi}
%
\end{equation}
 as $\chi$ increases.

In the case of white noise potential given in (\ref{f40}),
 the Lyapunov exponent
 has the form \cite{rmedia},
\begin{eqnarray}
 \lambda_{\omega} = \frac{\sqrt{\pi}}{2} N(\omega) \int_{0}^{\infty}
 \sqrt{y} e^{- y^{2}/12 - (R \omega / 2 D)^{2/3} y } dy    .
\end{eqnarray}
The entire curve is plotted in Fig.1.
%
%
We obtain the asymptotic behavior of the localization length
\begin{eqnarray}
  L_{loc}(\omega) \sim
\left\{
\begin{array}{ll}
 ( D / R \omega )^{2/3} & \mbox{for $ \omega \rightarrow 0 $}   \\
 1/D                   & \mbox{for $ \omega \rightarrow \infty $}      .
\end{array}
\right.             \label{f25}
\end{eqnarray}
%
%
%
%
%
We see that in the long wavelength limit $\omega \rightarrow 0$,
the wave function delocalizes.
This is a direct consequence of the multiplicative nature of the stochasticity
for classical wave propagation in random media \cite{WPRM}.

\section{Wave Propagation in Spacetimes with Time-Dependent
Metric Stochasticity:
 Parametric Amplification and Particle Creation}

Due to the conformal invariance of Maxwell's equations,
time-dependent fluctuations in the scale factor can be transformed away
(in a conformally-related coordinate). Assuming  spatial homogeneity
,the time dependence of the metric is uniform in space,
 we start from the metric with the conformal fluctuations  
similar to (\ref{f9}),
 but let the spatial part of the metric $f$ acquiring a time dependence.

If the background spacetime curvature is negligible,
the corresponding dielectric and peameability  functions take the
form
\begin{eqnarray}
	\epsilon_{ik} &=& \mu_{ik} = h(\eta) \delta_{ik}   \nonumber \\
			G_{i}   &=& 0
\end{eqnarray}
where we write the time-dependent factor as $h(\eta)$ instead of $f(\eta)$
for clarity.  Assuming $h(\eta)$ has the exponential form as in (\ref{f9})
\begin{equation}
	 h(\eta) = e^{ \nu(\eta)},
  \label{f14}
\end{equation}
%
then Maxwell's equations become
\begin{eqnarray}
 \nabla \times \vec{E} &=& - \frac{d}{d \eta}(h(\eta) \vec{H})  \nonumber \\
 \nabla \times \vec{H} &=& \frac{d}{d \eta}(h(\eta) \vec{E})   \nonumber \\
 \nabla \cdot \vec{E} &=& 0		   \nonumber \\
 \nabla \cdot \vec{H} &=& 0   .
\end{eqnarray}
From this we obtain a wave equation for $E$
\begin{eqnarray}
 \nabla \times \nabla \times \vec{E}
 = - \frac{d}{d \eta} (   h(\eta)
	  \frac{d}{d \eta}  	 h(\eta) \vec{E}	),
\end{eqnarray}
and a similar one for $\vec{H}$.

Since $\vec{E}$ has no divergence, for a slowly varying noise such that 
$ |\dot{h}(\eta)| << \omega h(\eta)$,
where $\omega$ is the frequency of $\vec{E}$,
 the above equation takes the simpler form
\begin{eqnarray}
 \frac{d^2}{d \eta^2}	\vec{E}
 - \frac{1}{h^2} \nabla^2 \vec{E} = 0
\end{eqnarray}
In momentum space, it reads
\begin{eqnarray}
	\frac{d^2}{d \eta^2}	\vec{E}_k
 + \frac{k^2}{h^2} \vec{E}_k    
 &=& \frac{d^2}{d \eta^2}	\vec{E}_k
 + k^2 e^{-2 \nu(\eta)} \vec{E}_k  = 0     .
\end{eqnarray}
Expanding in terms of $\nu$ gives
\begin{eqnarray}
 \frac{d^2}{d \eta^2}	\vec{E}_k
  + [k^2  -2 \nu(\eta)k^2] \vec{E}_k   = 0  .
	\label{f10}
\end{eqnarray}
This equation should be compared to (\ref{f6}) with $j = 0$
by identifying $V = 2 \nu k^2$ and $ E = k^2 $.

Following the argument from (\ref{f6}) to (\ref{f25}) and
noting that the time coordinate $\eta$ plays the role of space coordinate
$\chi$ here, we see
the qualitative features of  solutions to (\ref{f10}):
they show an exponential parametric amplification \cite{Zeldovich70}
with the rate characterized by the Lyapunov exponent $\lambda_h$
introduced in (\ref{f7})
\begin{equation}
 \lambda_h = \lim_{\eta \rightarrow \infty} \lambda_h(\eta)
= \frac{1}{2 L_{h}(k)}
\label{f53}
\end{equation}
and $\vec{E}_k$ grows
\begin{equation}
 \vec{E}_k \rightarrow \vec{E}_k(\eta = 0) e^{\lambda_h(\eta) \eta}
%
\end{equation}
 as $\eta$ increases.
 The characteristic time scale for parametric amplification
 $L_{h}(k)$ is equivalent to the localization length in
 (\ref{f7}) under the $\eta \rightarrow \chi$ correspondence.
 For a Gaussian white noise $\nu$ with correlation
 $
  \langle \nu(\eta) \nu(\eta')\rangle =
 D \delta(\eta - \eta'),
 $
the asymptotic behavior of $ L_{h}(k)$ is given by
\begin{eqnarray}
  L_{h}(k) \sim
\left\{
\begin{array}{ll}
  ( D / k )^{2/3} & \mbox{for $ k\rightarrow 0 $}   \\
  1/D  & \mbox{for $ k \rightarrow \infty $}  .
\end{array}
\right.             \label{f44}
\end{eqnarray}
%
%
%
%
%

It is  well-known that the cosmological particle creation problem in a  
FRW universe can be cast into an one-dimensional wave scattering 
 problem \cite{cpc}
by reading the time variable $\eta$ as a space variable,
and adopting the following boundary conditions:
\begin{eqnarray}
  \vec{E} \rightarrow
\left\{
\begin{array}{ll}
  e^{- i k \eta}     & \mbox{for $\eta \rightarrow -\infty $}   \nonumber \\
  \alpha e^{- i k \eta} + \beta e^{ i k \eta} 
                     & \mbox{for $\eta \rightarrow \infty $}     .
\end{array}
\right.
\label{f51}
\end{eqnarray}
In this analogy, the reflection coefficient $\beta$
  (or rather, $|\beta|^2$)in the wave scattering
picture gives the parametric amplification  factor and,
 in second quantization, the particle creation rate.
Since in this picture the transmission coefficient $t$ corresponds to the
incoming flux which is normalized to one, 
the effect of localization translates to an exponential increase
of the particle creation rate $\beta$ (so is $\alpha$)
\begin{equation}
 \beta  \rightarrow  e^{\lambda_h(\eta) \eta}   .
%
\end{equation}
More explicitly, if we compare (\ref{f51}) with the corresponding scattering
problem in which the wave packet is incident from $\eta = \infty $
and transit to $\eta = - \infty $
\begin{eqnarray}
  \vec{E} \rightarrow
\left\{
\begin{array}{ll}
  t e^{- i k \eta}     & \mbox{for $\eta \rightarrow -\infty $}  \nonumber \\
  e^{- i k \eta} + r e^{ i k \eta} 
		     & \mbox{for $\eta \rightarrow \infty $}
\end{array}
\right.
\label{f52}
\end{eqnarray}
we see that the particle creation rate $N = |\beta| = |r/t|^2$
in expanding universe
measures a dimensionless resistance in the wave transport picture.
The transmission coefficient in a random potential is a multiplicative quantity
 which decreases exponentially with length.
 This implies that the logarithm of the resistance is an additive quantity
 and its average in length becomes constant
  in the thermodynamic limit and hence non-fluctuating.
 The average of the resistance itself (hence the average particle creation
 rate) is also known to show the exponential increase in length
  ( in conformal time $\eta$ )
  with the rate characterized by $\lambda_h$.
We can obtain the asymptotic behavior of the average of $N$ in the high
energy limit as follows
\begin{eqnarray}
 \langle N \rangle = \langle |\beta|^2\rangle
	 = \frac{1}{2} ( e^{ 2 \eta / L_{h} } + 1 )    .
 \end{eqnarray}
Hence the fluctuations of $\langle N \rangle$ becomes
\begin{eqnarray}
\Delta N^2 = \langle N^2 \rangle - \langle N \rangle^2
	 = \frac{1}{8} e^{ 4 \eta / L_{h} }
		( e^{ 2 \eta / L_{h}} - 1 )
 \end{eqnarray}
for $ \eta >> L_{h}$ where $L_{h} = 1 / D$ is the characteristic particle
creation rate  which appeared in (\ref{f53}).
The exponential divergence of $\Delta N^2$ results in the
non-self-averaging nature of the observable.

\section{Hawking Radiation from a Fluctuating Black Hole}

In this section, we study the effect of metric stochasticity
in the Schwarzschild spacetime and
discuss  how it affects the Hawking radiation \cite{Hawking74}.

%
%
%

%
%
%

%
%
%

In the presence of metric stochasticity,
the Helmholtz equation acquires a fluctuating component in the refractive
index similar to the Friedmann-Robertson-Walker (FDR) universe case.
Assuming the same conformal form of stochasticity as in (\ref{f9}),
 but in a radial direction only for simplicity,
 the Helmholz equation in this case has the following form
\begin{equation}
	 \frac{d^2 \psi^{M}(r*)}{d r*^2}
	 + [ \omega^2 + 2 \sigma(r*) \omega^2 - U(r*) ]  \psi^{M}(r*) = 0
\end{equation}
where $\sigma(r*)$ is a stochastic variable as a function of $r*$.
If we assume the metric stochasticity is restricted
 in the finite region $-L < r* < 0$,
the incoming flux from future infinity suffers backscattering not only from
the potential $U(r*)$ but also from the randomness.
The absorption probability
$\Gamma_{\omega}$ is thus reduced by a factor of 
$e^{-\lambda_{\omega} L}$ where
$\lambda_{\omega}$ is the Lyapunov exponent which appeared in (\ref{f7})
with $R=1$.

%
%
%

Hence the total luminosity has the form
\begin{eqnarray}
 L = \frac{1}{2 \pi} \sum_{j=0}^{\infty} (2j+1) \int_{0}^{\infty}
 d\omega \omega \Gamma_{0} e^{-\lambda_{\omega} L} / (e^{8 \pi M \omega} -1).
\end{eqnarray}

%

%
%

 The more interesting case is when the fluctuating region contains
  the event horizon
 such that the stochasticity induces  horizon
 fluctuations \cite{Ford97,Sorkin97,Casher96}.
 For simplicity, we restrict the arguments to the scalar wave case
 in the rest of the section.
 The scalar wave in the metric (\ref{f26}) satisfies the equation
\begin{equation}
  \left\{
	 [
	 \rho
         r(\rho)( 1 - \frac{2M}{r(\rho)} )^{1/2}
	 \frac{ d }{ d \rho }
	 ]^2
	 +  \omega^2 r^4(\rho)
  \right\} \phi(\rho) = 0            .
  \label{f60}
\end{equation}
This takes on a simple form if we define the coordinate $x$ such that
$ dx = f(\rho) d \rho / \rho^2  $
\begin{equation}
  [ \frac{ d^2 }{ d x^2 }
	 +  \omega^2 r^4(x)
  ]  \phi(x) = 0              .
  \label{f61}
\end{equation}

If the metric in (\ref{f26}) has the conformal type of stochasticity
in the form $f_1(\rho) e^{2 \sigma (\rho)}$ and
 $f_2(\rho) e^{ \nu (\rho)}$,
 where $\sigma (\rho)$ and $\nu (\rho)$
 are two independent random variables with zero mean,
 the  equation corresponding to (\ref{f60}) becomes
\begin{equation}
  \left\{
	 [
   	 \rho r(\rho)( 1 - \frac{2M}{r(\rho)} )^{1/2} e^{ \sigma (\rho) 
         -  \nu (\rho) }
	 \frac{ d }{ d \rho }
	 ]^2
	 +  \omega^2 r^4(\rho)
  \right\}  \phi(\rho) = 0       .
  \label{f62}
\end{equation}
We can absorb $\sigma (\rho)$ by choosing the coordinate $x$ such that
$ dx = f(\rho)e^{ \nu(\rho) -  \sigma (\rho) } d \rho / \rho^2.  $
Then we have
\begin{equation}
  [ \frac{ d^2 }{ d x^2 }
	 +  \omega^2 r^4(x)  e^{ 4 \nu (\rho) }
  ]  \phi(x) = 0         .
  \label{f63}
\end{equation}
At the horizon, only incoming wave exists.
So the solution of (\ref{f63}) will be
\begin{equation}
\phi \sim \exp[ -i \omega r_s^2 e^{4\nu (\rho_s) } x  ]
  \label{f64}
\end{equation}
where  $r_s \equiv 2M$ and $\rho_s \equiv M/2$.
This mode function behaves as
\begin{eqnarray}
\phi &\rightarrow& 1 - i \omega r_s^2 e^{4\nu (\rho_s) } x
   \nonumber \\
	  &\rightarrow& 1 + \frac{ i \omega r_s^2 e^{4\nu (\rho_s) } }{\rho}
  \label{G}
\end{eqnarray}
as $x \rightarrow 1$ or, equivalently, $\rho \rightarrow \infty$.
Here we assumed the stochastic variables
 $\sigma (\rho)$ and $\nu (\rho)$
vanish at infinity.

Far from the horizon, the asymptotic flatness of the spacetime
allows us to write (\ref{f60}) as
\begin{equation}
  \left\{
	 [
         \rho r(\rho) \frac{ d }{ d \rho }
	 ]^2
	 +  \omega^2 r^4(\rho)
  \right\}  \phi(\rho) = 0
  \label{f66}
\end{equation}
 with the solution
\begin{eqnarray}
  \phi(\rho)
	 &=&
  \frac{1}{ \sqrt{\omega \rho} }
  [ A J_{1/2}(\omega \rho) + B J_{-1/2}(\omega \rho) ]
  \nonumber \\
	 &\rightarrow&
  \sqrt{ \frac{2}{\pi} }  \left[ A + \frac{B}{ \omega \rho } \right]
  \mbox{~~~~for $\omega \rho << 1 $}   .
  \label{E}
\end{eqnarray}
Comparing (\ref{E}) with (\ref{G}), we obtain the following 
absorption coefficient
for low energy scattering of the scalar wave,
\begin{eqnarray}
  \Gamma_a
	 &=&
  1 - \left| \frac{ 1 + i \frac{B}{A} }{ 1 - i \frac{B}{A} } \right|^2
  \nonumber \\
	 &\rightarrow&
  4 \omega^2 r_s^2 e^{4\nu (\rho_s) }
	 \mbox{~~~~for $\omega \rightarrow 0 $}     .
  \label{H}
\end{eqnarray}
From (\ref{H}), the absorption cross section for this process is
\begin{eqnarray}
  \sigma_a
	 &=&
  \frac{\pi}{\omega^2} \Gamma_a
  \nonumber \\
	 &\rightarrow&
	 4 \pi r_s^2 e^{4\nu (\rho_s) }
	 \mbox{~~~~for $\omega \rightarrow 0 $}
  \nonumber \\
	 &=&
	 4 \pi r_s^2 [ 1 + 4\nu (\rho_s)  + 8 \nu^2 (\rho_s) + \cdots ]  .
	\label{f69}
\end{eqnarray}
The average absorption cross section is then
\begin{eqnarray}
  \langle \sigma_a \rangle
	 &=&
  4 \pi r_s^2
  +  32 \pi r_s^2 \langle \nu^2 (\rho_s) \rangle
     \nonumber \\
	 &=&
  A_H
  + 8 A_H \langle \nu^2 (\rho_s) \rangle
	\label{f70}
\end{eqnarray}
 where $A_H \equiv  4 \pi r_s^2$ is the area of the event horizon
 and assume all the higher moments of $\nu$ vanish.
 It is known that low energy absorption cross section
 of the scalar wave
 gives the area of the event horizon
 for a large class of black hole solutions
 \cite{Unruh76}.

	Therefore, we expect the second term gives 
the horizon  area fluctuations
\begin{eqnarray}
 \delta A_H
 =
  8 A_H \langle \nu^2 (\rho_s) \rangle
	\label{f71}
\end{eqnarray}
 which induces fluctuations in the Hawking temperature
 \begin{eqnarray}
 \frac{ \delta T_{H} }{ T_{H} } =   4 \langle \nu^2 (\rho_s) \rangle .
	\label{f72}
\end{eqnarray}

%
%
%


\section{Discussion}

In this paper, we have studied the effect of metric fluctuations on 
wave propagation in flat and curved spacetimes.
We saw that for  electromagnetic waves  the effect of metric stochasticity
is equivalent to that of a random optical index.
With this analogy, we can calculate the correction to the scattering
cross section of the electromagnetic wave in curved spacetime with
 metric fluctuations.
 We also see the intensity fluctuations due to multiple scattering.
If we assume classical spacetime is effectively emerging
from the underlying quantum
fluctuations by coarse graining, it seems reasonable to assume that
the stochastic part reflects the symmetry of the background spacetime.
When the stochastic part of the metric has only radial dependence,
the problem is further reduced to an one dimensional transport
problem in which Anderson localization is manisfest.
For time-dependent but space-independent stochasticity, particle
creation rate shows exponential instability in conformal time.
Furthermore,
analogy with mesoscopic transport in one dimension predicts large
fluctuations in the particle creation rate due to its non-self-averaging
nature.
For Schwarzschild spacetime, 
time-independent stochasticity induces fluctuations
of the event horizon and ,correspondingly, Hawking temperature.

The restrictions we made on separate space and  time dependence of the
stochastic metric is mainly  for technical simplicity. 
We expect qualitatively similar phenomenon in the more general cases,
where space- and time -dependent metric fluctuations both exist. 
Certain modifications are expected such as dissipative effects on 
localization for a time-dependent random potential.

Possible origins of the metric fluctuations have been discussed recently.
Primordial stochastic gravitational waves is  currently under 
intense investigation \cite{Allen96,Ford95,Squeeze}. Stochasticity used in
such a context is different from that in stochastic semiclasical gravity
theory \cite{CH94,HM3,fdrsc,CamVer96}. 
In the former it refers to (quantized) linear perturbation 
of the metric obeying Einstein's equations whereas in the latter,
metric fluctuations is  induced by fluctuations in
quantum matter fields, which  obey  the Einstein-Langevin equation, as
a generalization of the semiclassical Einstein equation governing the mean
values of metric and matter stress energy. 
The naive replacement of the stress-energy tensor with its expectation value
leads to some pathological results such as the violation of the weak energy
condition of the Einstein equation \cite{KuoFor}.
Possible cure may be obtained by including the smearing term in the classical
background spacetime \cite{Wald}. 
Self-consistency between the classical gravitation 
and the quantum matter sectors also points to the inevitable dynamic role
 of metric and field fluctuations \cite{HuPhysica}. 
From the astrophysical viewpoint, cosmic string network may provide
the source of the stochastic
gravitational wave background which lies in the observable
frequency range of the LIGO detector  \cite{Allen96}.
The discussions and results in this work are more of a generic nature
insensitive to the origin of particular sources.

Ford argued that the lightcone fluctuations based on
linearized quantum gravity induce black hole horizon
fluctuations  \cite{Ford97}.
He suggests that the horizon fluctuations are small enough to validify
the semiclassical derivation of Hawking radiation.
Casher, et. al. argue that the semiclasscal approximation breaks down
close to but much larger than the Planck scale from the horizon
\cite{Casher96} due to the interaction with the atmosphere of the horizon.
 Sorkin's argument is based on Newtonian mechanics and
its vadility for general relativistic cases remains to be shown
\cite{Sorkin97}.
Since we assumed a static metric stochasticity
for the Schwarzschild spacetime, direct comparizon
with their results is not obvious.
Nevertheless, we believe our arguments have certain degrees of
generality,  knowing that the randomness can arise from the imhomogeneity
of a static spacetime.
These issues are currently under investigation.\\



%
%
%
%
%
%


\noindent {\bf Acknowledgement}
We thank Prof. P. Sheng for introducing us to the subject of wave propagation
in a random media, and Prof. E. Calzetta for useful comments. We enjoyed
the hospitality of the physics and 
mathematics department of the Hong Kong University of Science and Technology 
where part of this work was done.
 This work is supported in part by the  U S National Science Foundation
under grants PHY94-21849.

\noindent {\bf Figure 1}
The localization length is plotted as a function of frequency.
 R and D are set to 1.
\end{document}